\documentclass[journal,transmag]{IEEEtran}

\ifCLASSINFOpdf
\usepackage[pdftex]{graphicx}
\graphicspath{{img/jpeg/}}
\else
\fi
\begin{document}
\title{Magnetic Field of a Permanent Magnet }

\author{\IEEEauthorblockN{David Shulman\IEEEauthorrefmark{1}}

\IEEEauthorblockA{\IEEEauthorrefmark{1}Department of Chemical Engineering, Ariel University, Ariel, Israel 407000}

\thanks{
Corresponding author: D. Shulman (email: davidshu@ariel.ac.il)}}


\IEEEtitleabstractindextext{%
\begin{abstract}
For magnetic field calculations, cylindrical permanent magnets are often approximated as ideal, azimuthally symmetric solenoids. Despite the frequent usage of this approximation, research papers demonstrating the validity and limitations of this approach are scarcely available. In this paper, the experimentally derived magnetic field of a cylindrical permanent magnet is compared with the analytically calculated magnetic field of an ideal solenoid. An experimental setup for measuring the magnetic field distribution is demonstrated and employed for gathering the data. The proposed setup allows to measure the distributions of the axial and radial components of the magnetic field surrounding the magnet. The experimental data is in a very good agreement with the theoretical predictions, confirming the validity of using the model of an ideal solenoid for predicting a magnetic field distribution of a permanent magnet.

\end{abstract}

}

\maketitle
\IEEEdisplaynontitleabstractindextext
\IEEEpeerreviewmaketitle

\section{Introduction}
\IEEEPARstart{P}{ermanent} magnets are extensively used in a wide range of applications: electrical engineering, radio engineering, chemical engineering, medicine, household appliances and many others  \cite{myfirstarticle,TAI20085606,TREUTLER20012,sakamoto2007position,riley2002magnets} .Knowing and predicting magnetic fields of various electromagnetic configurations is crucially important for all relevant implementations. Already the early classical physicists were attempting to calculate electromagnetic fields of some widely-implemented configurations – finite helical solenoids, infinite solenoids and loops \cite{callaghan1960magnetic}. However, in those early studies only the simplest cases have been entirely calculated, such as the single loop for the latter configuration \cite{scott1959physics}. 
Currently several approaches to analytically define the magnetic field of a permanent magnet exist \cite{camacho2013alternative,ravaud2010cylindrical,selvaggi2010computing}. All these approaches are, however, very computationally expensive and usually make use of rather complex formulae or are valid only within certain limits. Just like the earlier works, modern analytical solutions focus on specific geometrical configurations – spheres, cylinders and other permanent magnets with azimuthal symmetry \cite{camacho2013alternative,dolisy2014three}, polyhedra \cite{rubeck2012analytical}, and even some more complex magnetic configurations, like multi-pole magnetic rings \cite{ausserlechner2012closed}. Therefore, many analytical solutions for the magnetic fields of various configurations have already been implemented and tested. M. Ortner and L. G. C. Bandeira \cite{ortner2020magpylib} have gathered these solutions and created a package allowing for magnetic field calculations of user-defined geometries. In our research we have used the above-mentioned software package for theoretical calculations of the magnetic field of an ideal solenoid. More details about the used formulae can be found in Section 2.
One of the most common approaches to testing a developed theoretical model for predicting a permanent magnet’s field topology is experimentally measuring the magnetic field distribution of a magnet of a certain structure and comparing it to the theoretical calculations of the same structure. This approach allows one to validate the developed theoretical model and discover the conditions under which the model is valid. Several theoretical approaches predicting the magnetic field distribution of a cylindrical permanent magnet currently exist: the magnet can be approximated as a point dipole \cite{hahn1998eddy,iniguez2004study}, a stack of several polygonal loops \cite{maclatchy1993quantitative}, or an ideal solenoid \cite{derby2010cylindrical}. An ideal solenoid model is used as an approximation for calculating the magnetic field of a cylindrical permanent magnet the most frequently \cite{kwon2020magnetic,ravaud2010cylindrical}. However, detailed comparisons of experimental measurements with theoretical calculations are difficult to find. The most notable comparison was performed in \cite{derby2010cylindrical}, where the authors have validated the ideal-solenoid model by comparing its results with the measurements of the well-known experiment of dropping a cylindrical permanent magnet into a vertical non-magnetic tube of a certain conductivity. During the experiments the total time of fall of cylindrical magnets through a copper tube was measured and subsequently compared with the calculated values. The authors have demonstrated a very precise agreement between theory and experiment.  
In this paper we demonstrate a straight-forward experimental approach of measuring the distributions of the axial and radial components of the magnetic field surrounding a permanent magnet. We then proceed to compare the obtained experimental results with the predictions of an ideal solenoid model.

\section{Theoretical basis of the suggested experimental method}
\label{section:theor}
The magnetic field of an ideal solenoid can be computed from the Biot-Savart law. The formulae used for the calculations are presented below: the $ B_r\ (r,h)$ represents the radial component of the field, and the $ B_z\ (r,h)$ – the longitudinal, or axial, component of the field. The exact derivation of the formulae can be found in \cite{derby2010cylindrical}. To develop the formulae, Derby et al. have divided the surface of the solenoid into circular stripes of equal width. They have then defined the magnetic field at a given point in space as the sum of the magnetic fields from each loop; where the magnetic fields of each loop were calculated with the Biot-Savart law. 

\begin{figure*}[!t]
\normalsize
\begin{equation}
\label{eqn_dbl_x}
B_r\ (r,h)=B_0 \int_0^{\pi/2}d\psi(cos^2\psi-sin^2\psi)\left( \frac{\alpha_+}{\sqrt{cos^2 \psi+k_+^2 sin^2 \psi}} -\frac{\alpha_-}{\sqrt{cos^2 \psi+k_-^2 sin^2 \psi}})\right)
\end{equation}
\begin{equation}
\label{eqn_dbl_y}
B_z\ (r,h)=B_0\frac{a}{r+a}\ \int_0^{\pi/2}d\psi \frac{cos^2\psi+\tau sin^2\psi}{cos^2\psi+\tau^2 sin^2\psi} \left( \frac{\beta_+}{\sqrt{cos^2\psi+k_+^2 sin^2\psi}} -\frac{\beta_-}{\sqrt{cos^2\psi+k_-^2 sin^2\psi}})\right)
\end{equation}

\hrulefill
\vspace*{4pt}
\end{figure*}

$$\alpha_\pm=\frac{a}{\sqrt{h_\pm^2+(r+a)^2\ }}$$
$$\beta_\pm=\frac{h_\pm}{\sqrt{h_\pm^2+\left(r+a\right)^2}}$$
$$h_-=h-2b,\qquad h_+=h$$
$$\tau=\frac{a-r}{a+r}$$
$$k_\pm=\sqrt{\frac{h_\pm^2+\left(a-r\right)^2}{h_\pm^2+\left(a+r\right)^2}}$$
$$B_0=\frac{\mu_0}{\pi}nI,$$

where a is the radius and 2b is the length of the solenoid; $\left(r,\ \varphi,\ h\right)$ are the cylindrical coordinates with the origin at the center of the solenoid; n – is the number of turns per unit length. To obtain the equations in the current form, they have also introduced the following integration variable change: $2\psi\equiv\pi-\varphi^\prime$. To compare the calculation results with the results of the measurements, the radius and the length of the solenoid for the calculations were chosen equal to the radius and the length of the permanent magnet investigated experimentally. 
The above-mentioned equations were used for the calculations of the axial and radial components of the magnetic field of the ideal solenoid. The calculations were performed using Magpylib – a free Python package for magnetic field computation \cite{ortner2020magpylib}. 

\section{Experimental arrangement}
\begin{figure*}[!t]

\normalsize
  \begin{minipage}{0.48\textwidth}
     \centering
     \includegraphics[width=.7\linewidth]{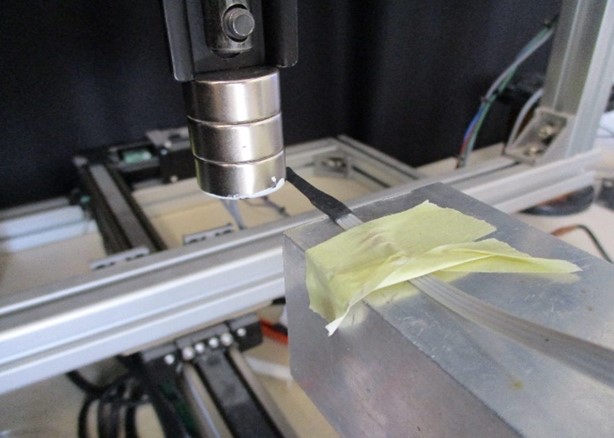}
     \caption{The photograph of the experimental setup}\label{Fig:Data1}
  \end{minipage}\hfill
  \begin{minipage}{0.48\textwidth}
     \centering
     \includegraphics[width=.7\linewidth]{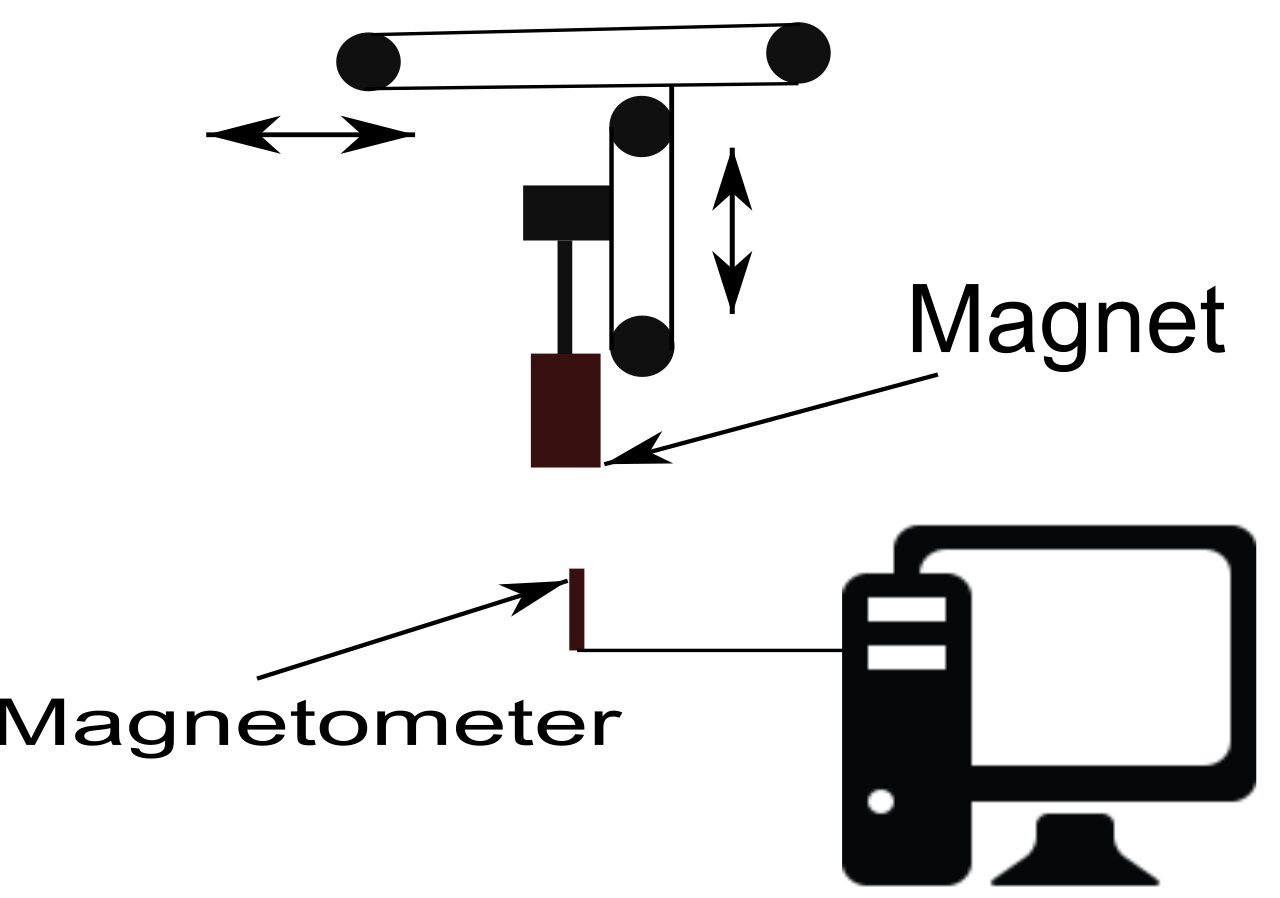}
     \caption{The schematic outline of the experiment setup}\label{Fig:Data2}
  \end{minipage}
\hrulefill
\vspace*{4pt}  
  \begin{minipage}{0.48\textwidth}
     \centering
\includegraphics[width=\linewidth]{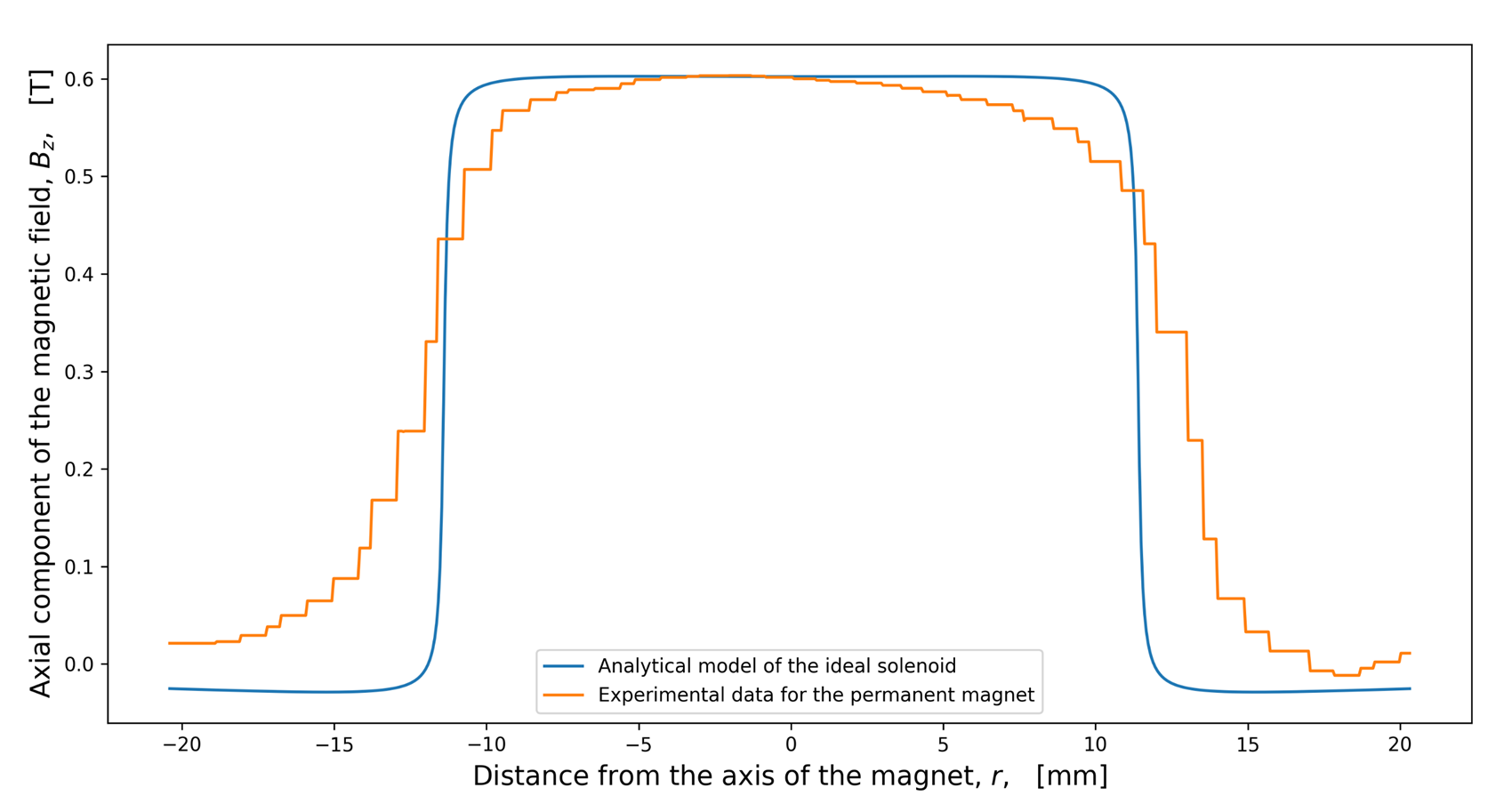}
\caption{Axial component of the magnetic field of magnet 1. The distance of the magnet from the magnetometer in the experiments h=0.1\ mm. The coefficient of determination $R^2=0.867$.}
\label{fig:Axial_big_One}
  \end{minipage}\hfill
  \begin{minipage}{0.48\textwidth}
     \centering
\includegraphics[width=\linewidth]{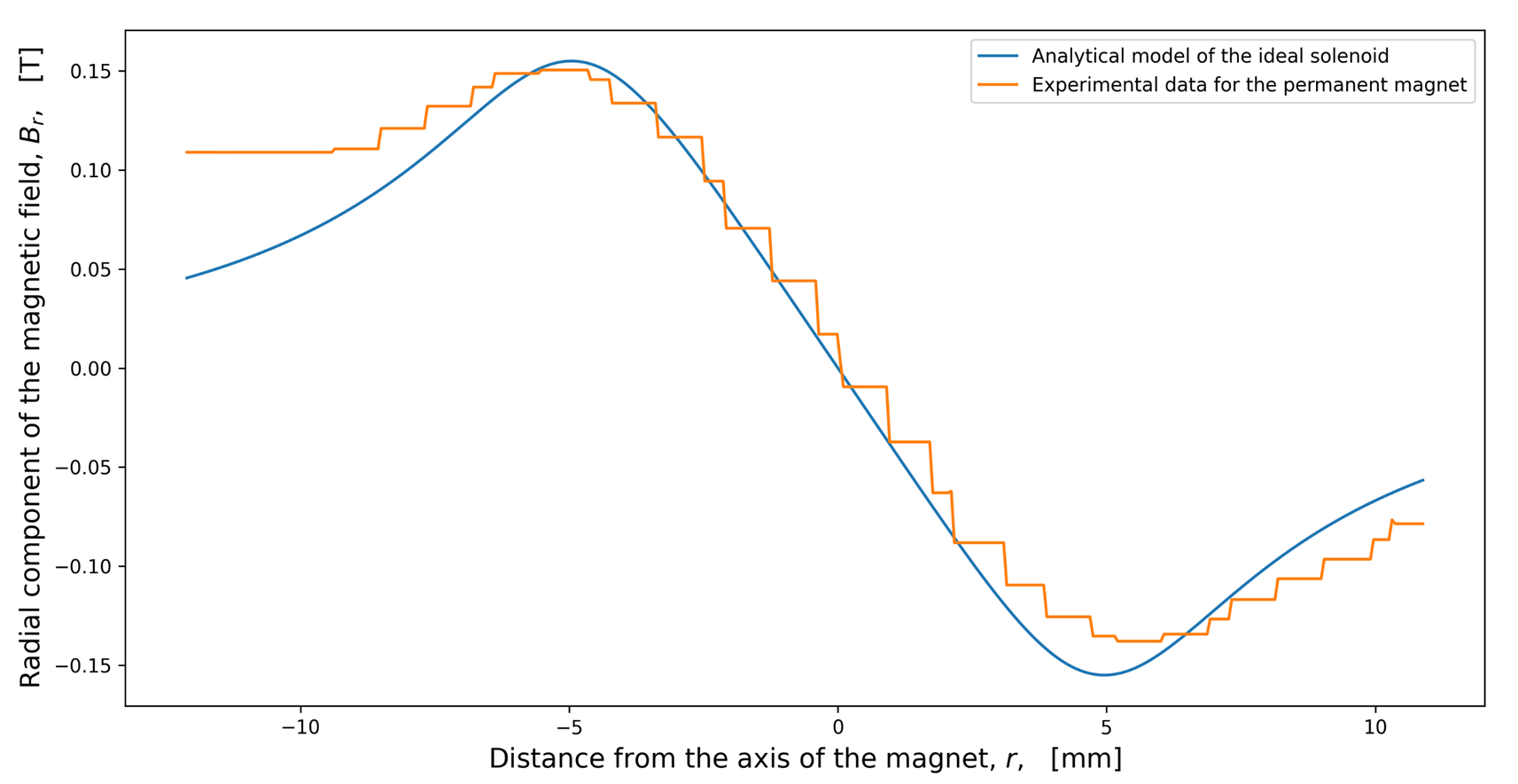}
\caption{Radial component of the magnetic field of sample 2 (10 mm x 27 mm). The distance of the magnet from the magnetometer in the experiments h=2.5 mm. The coefficient of determination $R^2=0.95$. }
\label{fig:Radial_small_one}
  \end{minipage}
\vspace*{4pt} 
\end{figure*}

The experimental setup used for measuring the axial and radial distributions of the magnetic field of a permanent magnet is demonstrated in figures \ref{Fig:Data1}-\ref{Fig:Data2}, where figure \ref{Fig:Data1} presents the picture of the actual setup, and figure \ref{Fig:Data2} – the basic outline of the setup. The main element of the setup and the object under investigation was a stack of cylindrical neodymium (NdFeB) magnets. Two stacks of magnets with different dimensions were used during the investigations: the first sample had a diameter of 22.8\ mm and a length of 30\ mm (where each magnet in a stack had a length of 10\ mm); the second sample had a diameter of 10\ mm and the length of 27\ mm (where each magnet in a stack had a length of 9\ mm). The magnetic field strength at the top surface of each stack at the axis of the magnet, or the maximum magnetic field, was measured prior to the experiments and turned out to be $B_z \approx 0.6 \pm0.01\ T$ and $0.55 \pm0.01\ T$ respectively. The initial data for both samples is summarized in Table \ref{table:1}. 

\begin{table}[!t]
\caption{Parameters of the neodymium magnetic samples used in the experiments.}
\label{table:1}
\centering
\begin{tabular}{||c|c|c|c|c||} 
 \hline
 Sample nr. & Diameter, [mm] & Length, [mm] & $B_0$, [T] & $B_{z,\ max}$, [T] \\ [0.5ex] 
 \hline\hline
  1 & 22.8 & 30 & 1.3 &0.6 \\
 2 & 10 & 27 & 1.15 &0.55\\ [1ex] 
 \hline
\end{tabular}

\end{table}

The maximum magnetic field, as well as all values of the magnetic fields during measurements were obtained with a direct current/alternating current GM2 Gauss Meter, manufactured by AlphaLab Inc., USA (with an accuracy of $\pm0.01\ T$). The magnetometer mentioned above is not equipped with a direct connection to a PC, therefore, a special procedure was developed to record the measured values of the magnetic field. The procedure was as follows: the magnetometer was set to continuously measure the magnetic field of the sample under investigation, and a separate camera was installed directly facing the screen of the magnetometer. The video recorded by the camera was saved, and subsequently analyzed. The analysis included dividing the video sequence onto separate frames, and analytically determining the values displayed on the screen. These values were saved into a separate file, which was used for plotting the data.
To control the location of the sample, an XYZ actuator provided by CCM Automation Technology was modified and employed to move the magnet above the magnetometer, both in the vertical and in the horizontal directions. The measurement was performed as follows: as the first step, the vertical position of the permanent magnet was fixed at a certain value, starting with the magnet right above the surface of the test probe of the magnetometer (at a distance of $0.1\pm0.05\ mm$ for sample 1 and $0.3\pm0.05\ mm$ for sample 2). Consequently, the XYZ actuator initiated the constant movement of the sample above the magnetometer. The horizontal speed of the magnet was $v=1.42\pm0.02\ \frac{mm}{s}$. On average, the magnetometer makes 2 measurements per minute. Taking into account the speed mentioned above, the measurements were performed approximately every $1.57\ mm$. The data measured by the magnetometer was saved with the frame rate of about 2 frames per second. Then the magnet was lifted vertically with the step of $0.5\pm0.05\ mm$ and the measurement process was repeated. The data measured by the magnetometer was saved with the frame rate of 29.85 frames per second. In principle, the magnetometer only measures the axial component of the magnetic field. To measure the radial component, the test probe of the magnetometer was rotated $90\pm2$ degrees, after which the measurements were again repeated for both samples.

\section{Results and discussion}
As discussed in the above section, two cylindrical magnetic samples with different dimensions have been used during the experimental procedures. We will first discuss the results concerning the axial component of the magnetic field for both samples.
Figure \ref{fig:Axial_big_One} demonstrates one set of measurements with sample 1 fixed at the distance of 0.1 mm above the test probe of the magnetometer, combined with the calculated results of the analytical model described in Section \ref{section:theor}. The coefficient of determination in this case was $R^2=0.867$. Similar graphs were obtained for multiple vertical positions of each sample. For sample 1, the vertical distance of the magnet from the magnetometer was varied from 0.1 mm to 9.6 mm. For sample 2 this distance varied from 0.3 mm to 9.8 mm.

\begin{figure*}[!t]
\normalsize
\includegraphics[width=\linewidth]{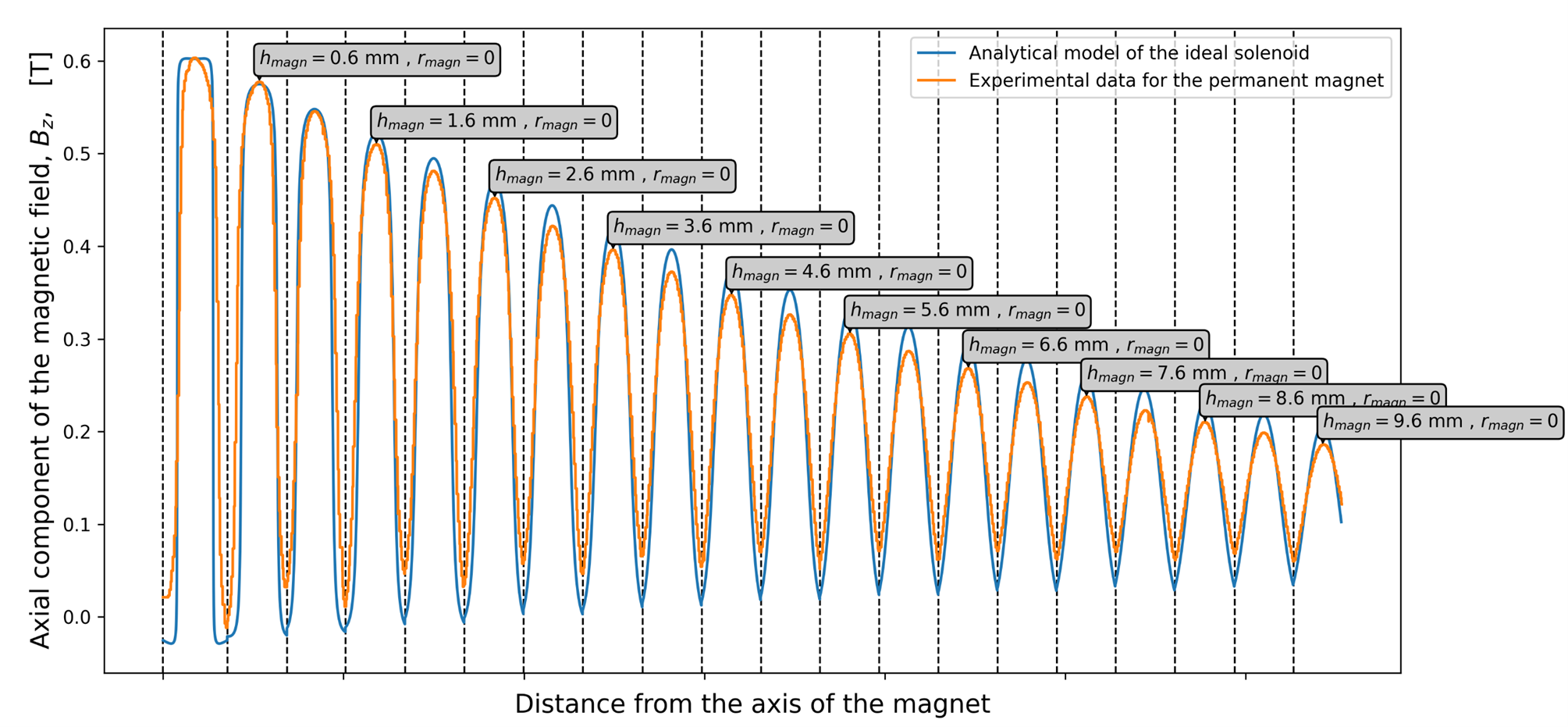}
\caption{The overall plot of the axial component of the magnetic field of sample 1 (22.8 mm x 30 mm). The axial distance from the sample probe of the magnetometer varied from 0.1 to 9.6 mm. Vertical lines represent the borders of the subsequent plots of different vertical positions of the magnet.}
\label{fig:Axial_big_Many}
\includegraphics[width=\linewidth]{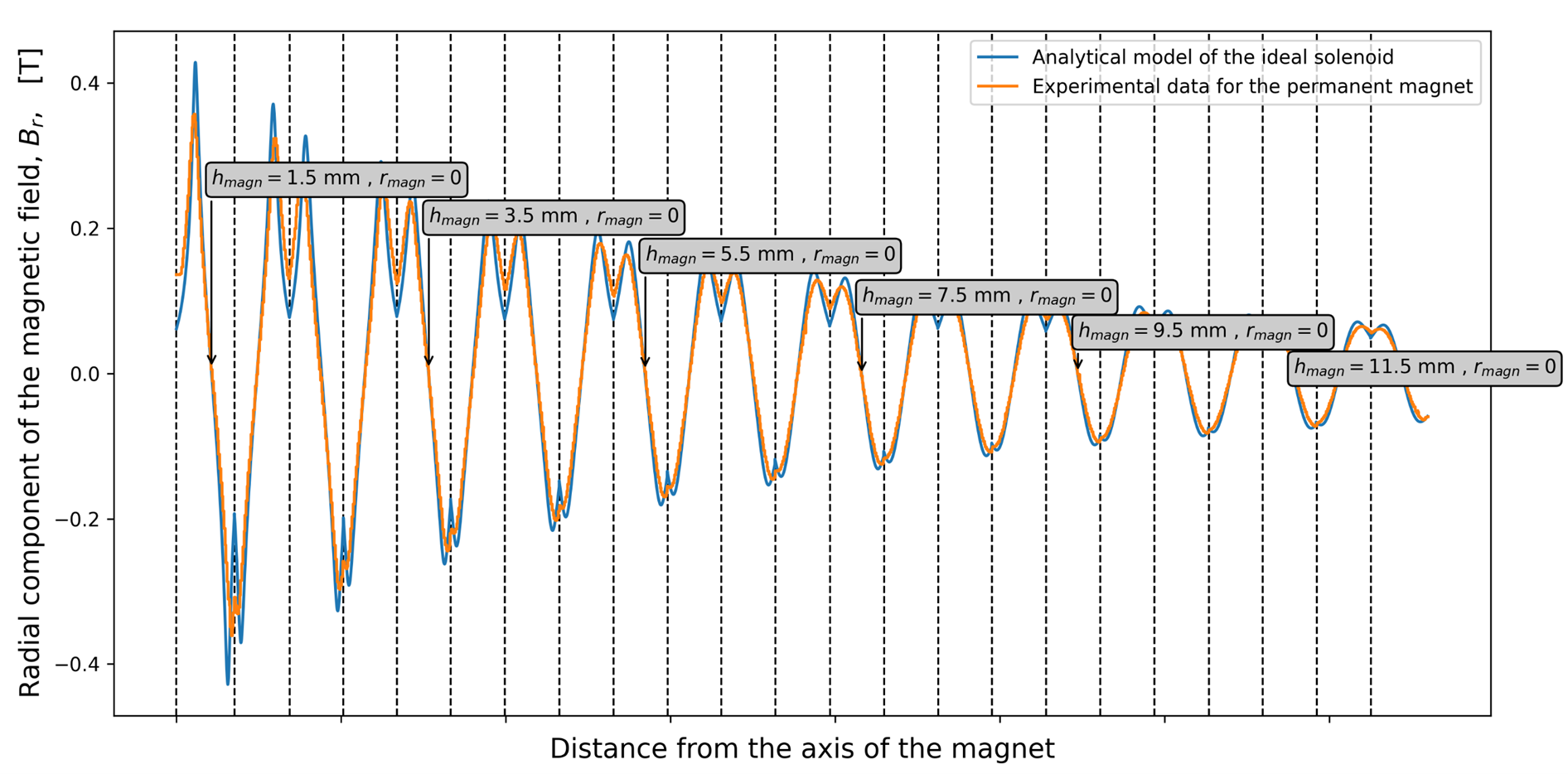}
\caption{The overall plot of the radial component of the magnetic field of sample 1 (22.8 mm x 30 mm). The distance from the sample probe of the magnetometer varied from 1.5 to 12.5 mm.  Vertical lines represent the borders of the subsequent plots of different vertical positions of the magnet. }
\label{fig:Radial_big_many}
\end{figure*}

\begin{figure}[htbp]
\centering

\includegraphics[width=\linewidth]{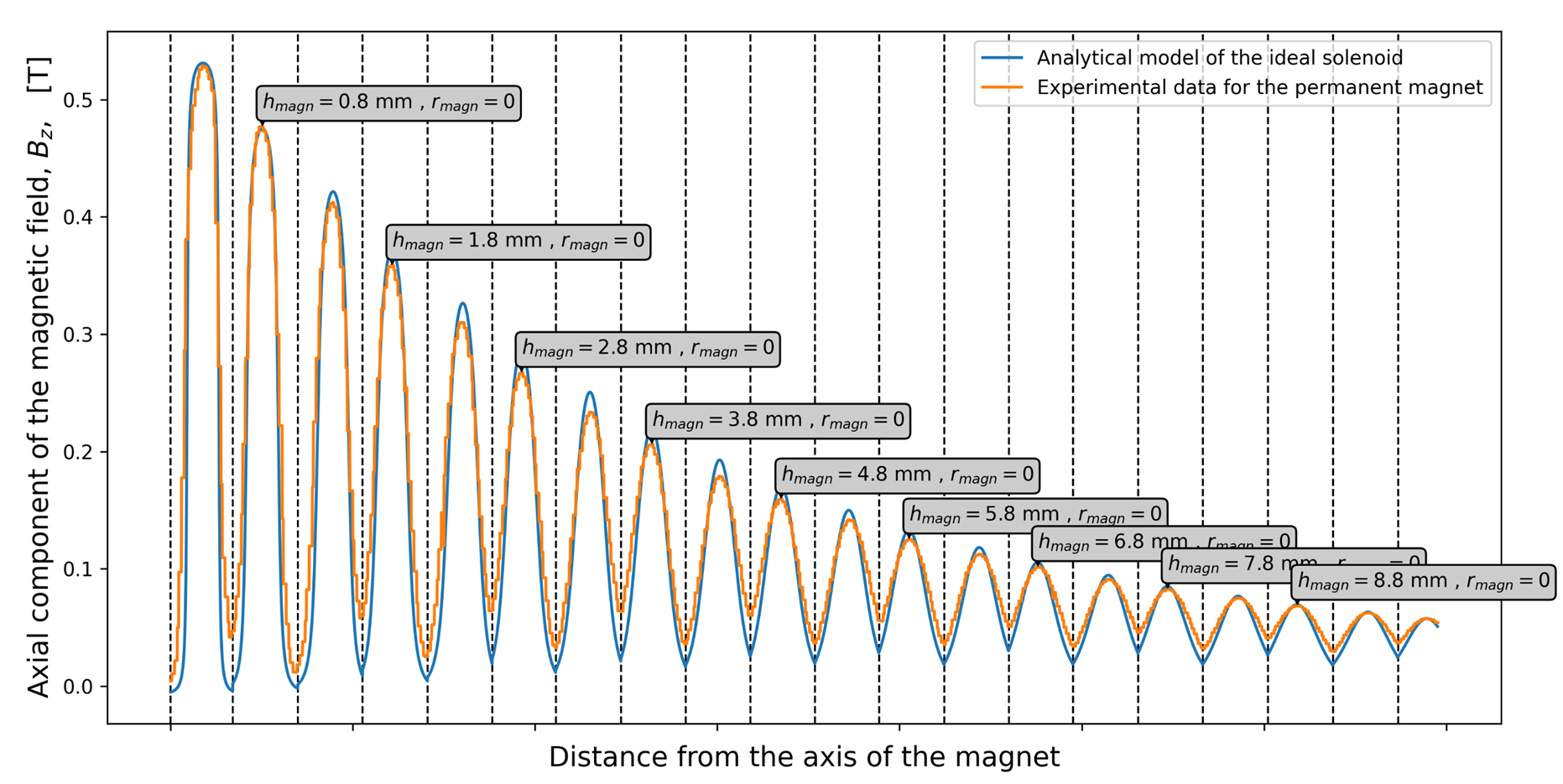}
\caption{The overall plot of the axial component of the magnetic field of sample 1 (22.8 mm x 30 mm). The axial distance from the sample probe of the magnetometer varied from 0.1 to 9.6 mm. Vertical lines represent the borders of the subsequent plots of different vertical positions of the magnet.}
\label{fig:Axial_small_Many}
\end{figure}

The measurements at each fixed vertical position for magnet 1 were performed up to the horizontal distance of 20 mm from the central axis of the magnet, on each side of the magnet (therefore, the distance from the axis on figure 1 varies from -20 mm to +20 mm). For better analysis, the measured and calculated values for various horizontal and vertical positions of the magnet were combined in one plot, where, as soon as the measurement data at +20 mm distance was plotted, a subsequent plot of the next vertical position starting with the horizontal distance of -20 mm was attached to it. The resulting overall plot for the axial component of the magnetic field for sample 1 is presented in Figure \ref{fig:Axial_big_Many}. Figure \ref{fig:Axial_small_Many} presents a similar graph for sample 2. For this sample the measurements were performed up to the horizontal distance of 10 mm  from the central axis of the magnet.

The analytically calculated values appear to be in a very good agreement with the measurements of the axial component of the magnetic field of two cylindrical permanent magnets of different dimensions.

The following part of the section focuses on the radial component of the magnetic field. As stated in section 3, to perform experimental measurements of the radial component, the sample probe of the magnetometer was rotated 90°. Figure \ref{fig:Radial_small_one} presents a single measurement of the radial component of the magnetic field for sample 2 fixed at a distance of 2.5 mm from the magnetometer, along with the analytically calculated values. The coefficient of determination here was $R^2=0.95$. Again, similar figures have been plotted for each vertical position of both samples.

\begin{figure}[htbp]
\centering
\includegraphics[width=\linewidth]{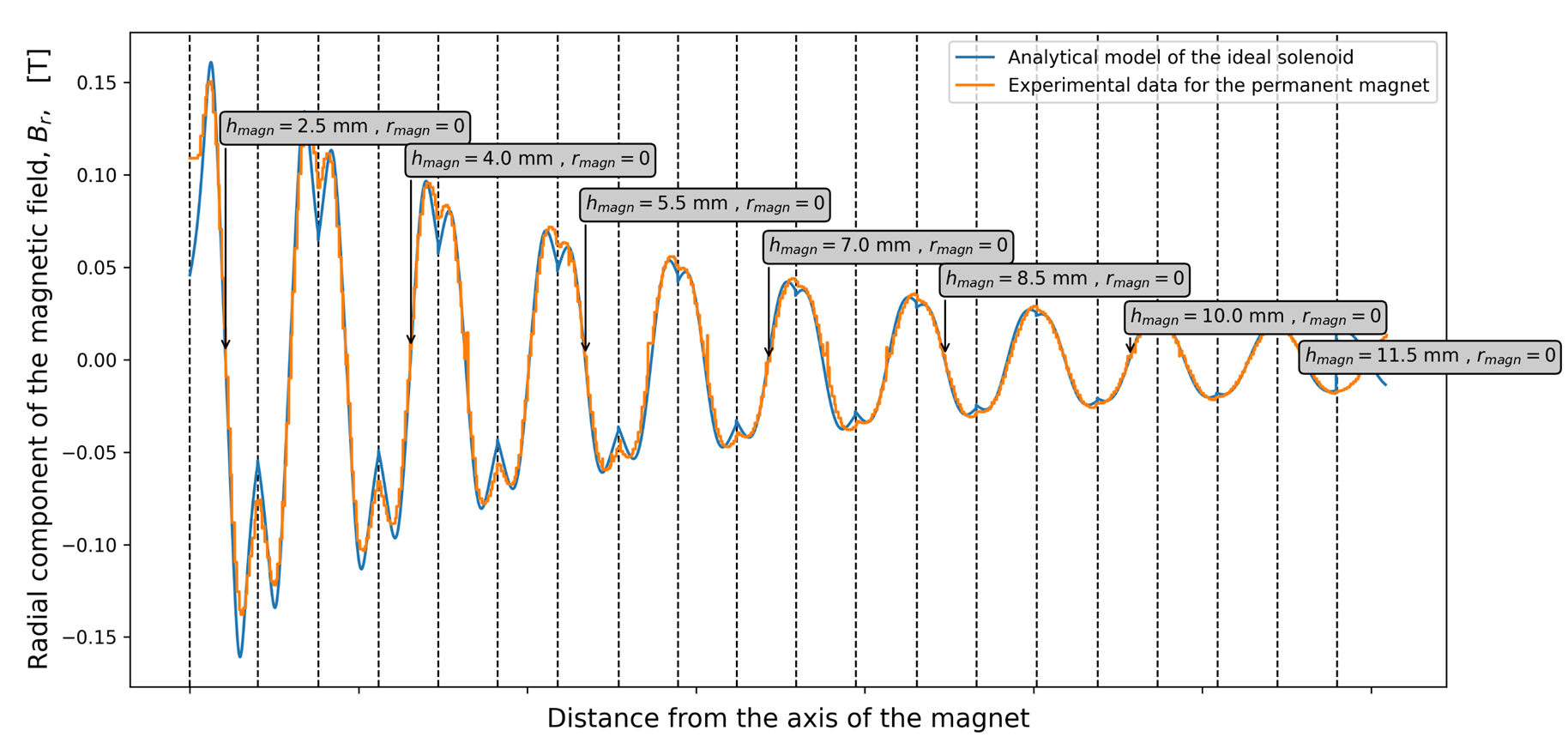}
\caption{The overall plot of the radial component of the magnetic field of sample 2 (10 mm x 27 mm). The distance from the sample probe of the magnetometer varied from 2.5 to 11.5 mm . Vertical lines represent the borders of the subsequent plots of different vertical positions of the magnet.}
\label{fig:Radial_small_many}
\end{figure}

Similar to the measurements of the axial component, the radial component was measured up to a certain distance from the axis of the magnet on both sides. During these measurements, the magnetic field values were recorded up to a distance of 25 mm on both sides of the magnet for sample 1, and 12 mm for sample 2 . Furthermore, the compound graphs of the radial component of the magnetic field for all positions of the samples have been plotted. The resulting graphs for samples 1 and 2 are presented in figures \ref{fig:Radial_big_many} and \ref{fig:Radial_small_many} respectively.


Also here, the results of the experimental measurements are in an extremely good agreement with the analytical calculations. 

Experimentally measured values of the axial and the radial components of the magnetic field of a cylindrical permanent magnet demonstrated here, appear to be in a perfect agreement with the theoretical predictions based on the model of an ideal solenoid. 
These results provide a strong confirmation of the validity of employing the model of an ideal solenoid for analytical calculations of the magnetic field generated by a cylindrical permanent magnet.

\pagebreak
\section{Conclusions}
Both the axial and the radial components of the magnetic field of several cylindrical permanent magnets have been measured and compared with the calculations obtained with the help of the model of an ideal solenoid. The theoretical values turned out to be in a very good agreement with the experimental data. Therefore, this paper demonstrates a long-required confirmation of the validity of using the analytical model of an ideal solenoid to predict and describe the magnetic field of a cylindrical permanent magnet.

\section*{Acknowledgment}

I would like to express my deep gratitude to Professor Meir Lewkowicz and Professor Edward Bormashenko, my research supervisors, for their patient guidance, enthusiastic encouragement, and useful critiques
\section*{AUTHOR DECLARATIONS}
\subsection*{Conflict of Interest}
The author has no conflicts to disclose.

\newpage
\ifCLASSOPTIONcaptionsoff
  \newpage
\fi

\bibliographystyle{IEEEtran}
\bibliography{mybibfile}
\end{document}